TITLE: Mars Reconnaissance Orbiter's Mars Color Imager (MARCI): A New Workflow for Processing Its Image Data

RUNNING HEAD: Processing MRO's MARCI Data

Stuart J. Robbins*,[a] ORCID: 0000-0002-8585-2549

*stuart@boulder.swri.edu
*Corresponding Author
[a]Southwest Research Institute, 1050 Walnut St., Suite 300, Boulder, CO 80302, United States

Total Pages – 39; Total Tables – 0; Total Figures – 7

Key Points:

- A discussion of how to process MARCI images is presented with discussion of issues due to camera construction and software limitations.
- Mosaics and composites for different time durations are shown, showing the utility of each for different scientific investigations.
- A small area of Valles Marineris illustrates the detail afforded by MARCI for monitoring changes on Mars at sub-kilometer scales.

Plain Language Summary (200-word limit):

The first modern spacecraft to orbit Mars was NASA's *Mars Global Surveyor*, which had the Mars Orbiter Camera that – among other imaging capabilities – provided global, daily images at about 7 kilometers per pixel from 1999 through 2006. Its successor is the Mars Color Imager, or MARCI, onboard NASA's *Mars Reconnaissance Orbiter*, which has been taking images since 2006. MARCI builds daily, global images by imaging a swath of Mars at up to 0.5 kilometers per pixel, once per orbit, and it orbits Mars about 13 times per day. These swaths are taken in seven different colors ranging from ultraviolet to infrared. Because of how MARCI was built and how its images are taken, processing its images into both a scientifically and visually useful product is difficult. Thus, MARCI data are not widely used. This paper presents a new workflow built on open tools to process the data, discussion on how to get around some processing difficulties, and illustrates the workflow's success by showing several examples of how MARCI can be used to show a dynamic world of huge cloud systems, small dust storms, and surface changes.

Abstract (250-word limit):

The *Mars Reconnaissance Orbiter*'s (*MRO*'s) Mars Color Imager (MARCI) has returned approximately daily, approximately global image data of Mars since late 2006, in up to seven different colors, from ultraviolet through near-infrared. To-date, that is over 5300 Mars days of data, nearly eight full Mars Years, or more than 15 Earth years. The data are taken at up to nearly 500 meters per pixel, and the nearly circular orbit of *MRO* and its consistent early afternoon imaging provide an unprecedented baseline of data with which to study Mars' atmosphere and surface processes. Unfortunately, processing MARCI data is difficult, fraught with exploding file sizes, issues that require workarounds in free software, and other problems that make this a severely under-utilized dataset. This paper discusses a workflow to process MARCI data to their fullest, including suggestions on how to work around issues unique to MARCI and how the data work with the current version of the free software *ISIS* (*Integrated Software for Imagers and Spectrometers*). Discussion of some trades that can be made to dramatically speed data processing are also described. Examples of processed MARCI images, mosaics, and color composites are shown, demonstrating the abilities of this workflow on global, regional, and local areas at the full, 96 pixels per degree scale afforded by MARCI.

Keywords: Mars; MARCI; global data; data processing; mosaics

## 1. Introduction and Context

Earth-based monitoring of Mars has a long history, with researchers searching for any changes to its fuzzy, rusty colored disk. *Viking Orbiter 1* and *2* presented the first orbital monitoring, orbiting Mars for a few years and providing well resolved observations of changing ice and frost, weather patterns, and other features not visible from Earth-based observatories. A gap in coverage lasted more than 20 years, and NASA returned to Mars in 1997, with the *Mars Global Surveyor* (*MGS*). *MGS* lived up to its name and provided the first in a series of globe-spanning observations from different spacecraft that, to this day, provide better-than-daily coverage of Mars from aerocentric orbits.

Aboard *MGS* was the Mars Orbiter Camera, and its wide-angle camera (MOC-WA; Malin et al., 1991) in high spatial summing modes allowed it to take blue and red images of swaths of Mars with a cadence slightly better than every two hours. Those swaths built a nearly global view, every Mars day, at an approximately consistent time (early afternoon) and consistent pixel scales of ~6.5 km/pix. Four years after *MGS*'s science mission began, ESA's *Mars Express* (*MEx*) arrived, providing images with its engineering VIsual monitoring Camera (VIC; Ormston et al., 2011; Sánchez-Lavega et al., 2018) and science High-Resolution Stereo Camera (HRSC; Jaumann et al., 2007). VIC is a low-resolution RGB (red-green-blue) camera and takes images at apoapsis, and HRSC is a blue, green, far-red, and near-IR (~1 μm) camera. ISRO's *Mars Orbiter Mission* (*MOM*) has the Mars Color Camera (MCC; Arya et al., 2015), an RGB camera similar to VIC which, can take full-disk images at apoapsis, and at periapsis it can take higher resolution images at the several-hundred-meters scale. MCC started imaging Mars in 2013, overlapping with VIC and HRSC. Both *MEx* and *MOM* have highly elliptical orbits that do not replicate ground coverage well, and the elliptical orbits present an inconsistency in pixel scale within images, making it difficult to generate a consistent product. However, these orbits mean that their cameras have the ability to provide a range of time-of-day imaging not afforded to NASA craft that image at consistent times of day.

Overlapping with *MGS* for just a few weeks in 2006 was the successor to MOC-WA, the Mars Color Imager (MARCI; Malin et al., 2001; Bell et al., 2009) aboard the *Mars Reconnaissance Orbiter* (*MRO*). *MRO* is in a fairly circular orbit around Mars, with consistent early afternoon and early after-midnight imaging at nadir pointing. MARCI images an approximately longitudinal swath roughly 30° wide, building up a nearly global mosaic every Mars day if images are taken each orbit. MARCI operates in two ultraviolet (UV) bands and five visible / near-IR (VIS) bands, with effective central passbands from 260 nm to 720 nm, sensitive at the tails from about 230 nm through 780 nm (with some gaps). It is capable of taking images at roughly 600 m/pix in VIS and 4800 m/pix in UV from *MRO*'s orbit.

While not the highest pixel scale camera, MARCI is the best camera for stable, consistent observations on a roughly daily basis that currently exists, and its very long baseline of consistent observing (2006 to present) spans over 5300 Mars days. However, MARCI data are very rarely used in contrast with other science cameras on recent spacecraft. While not everyone cites instrument papers, and Cornell's ADS cannot track all citations, one can still use that tool to get a relative idea of use. For example, McEwen et al. (2007) for the High-Resolution Imaging Science Experiment on *MRO* has nearly 800 citations to-date, and HRSC has over 250 citations to-date. In contrast, the two instrument papers for MARCI have 48 (Malin et al., 2001) and 43 (Bell et al., 2009). Those are relative rates of ≈55 and ≈15 per year, versus 2.4 and 3.5 per year, literally an order-of-magnitude difference.

While it is true that high-resolution images are perhaps more interesting to researchers, this is still a marked difference. It is my opinion that one of the primary impediments to using MARCI is that it is difficult to process for various reasons discussed later in this text. The few researchers who use MARCI have developed their own, proprietary methods to process it, often outside of free software. Unfortunately, my own institution considers the workflow I developed to be similarly proprietary, so I instead describe in detail a method I developed to process the MARCI data, all with open-source software. This includes a discussion of several issues that arose during development, some of which might be solved in the near future by the developers of that open-source software. Appendix A includes a working set of commands to process a MARCI image.

Section 2 of this work describes why MARCI data are difficult to use, and how those difficulties can, at least in part, be mitigated. Section 3 describes in detail the methods to process the MARCI data. Section 4 shows some sample data and for what MARCI can be used, including a very brief literature review of how others have used these data. Section 5 provides a brief summary and discussion. Appendix B describes some quirks or issues with the free processing software used, while Appendix C and D provide further discussion of two more MARCI processing quirks.

## 2. Why MARCI Data Are Hard to Use

Successful processing of MARCI data into a usable product requires mitigating several issues based on how MARCI was built and the *MRO* imaging campaign.

First, *MRO* is in a science orbit ≈250–316 km above Mars' surface, whereas *MGS* was ≈373–437 km. Therefore, to fully image Mars with approximately overlapping swaths once per day, MARCI's field of view must be larger: 180° for MARCI versus 140° for MOC. Such a wide field presents significant optical distortions even just a quarter of the detector away from the center. Even with a good camera model, the viewing geometry results in high emission angles, atmospheric blurring, and atmospheric extinction that must be mitigated (or, those data simply discarded).

Second, MARCI is a push-frame camera, only the third that NASA has flown to other bodies. The push-frame means that instead of a classic framing (a 2-D array of pixels that are exposed simultaneously) or a push-broom (usually a 1-D array of pixels that scans across the surface to build an image), the two concepts are merged. MARCI has two optical paths, split into VIS and UV, that expose a CCD with filters bonded to it. The VIS's filters are five, each covering 1024×16 pixels, while the UV has two filters and the same number of pixels each, but it is always operated in an 8×8 spatial summing mode, reducing the detector to effectively 128×2 pixels. Each frame (all colors) is exposed as individual framelets (each filter, one at a time) in a carefully-timed manner over the course of nearly an hour so the spacecraft's motion builds up a full, seven-color image with overlapping framelets for each color so there are no gaps (though this was not perfected until about the sixth month of imaging). There are two primary difficulties that come from this imaging method: Multiple pixels at the same location, and large file sizes. The former is a problem because the free, United States Geological Survey (USGS) software *ISIS* (*Imaging Software for Imagers and Spectrometers*; Anderson & Sides, 2004) cannot handle multiple pixels in a single image with the same latitude and longitude location (or, back-tracking to cover some of the same terrain in a single image), while the latter arises because each color should be split into different files to process, and each has different geometric data since they are exposed at different times. *ISIS*'s method to handle the first problem is that images are split into even/odd framelets, per color, meaning that one swath of Mars becomes 14 images to process separately. The math is also done in 32-bits, doubling file sizes. Due to some optimization of the .cub format, a 5-color image is increased on-disk by a factor of 8, so one VIS image swath might be 110 MB downloaded from NASA's Planetary Data System (PDS), and that will become ≈880 MB to process on-disk; several *ISIS* programs write to separate files, meaning this space must be at least doubled again. While ~2 GB might not seem like a lot given today's common data storage devices, there are typically ~300 VIS images for one Earth month.

Third, every material has a "phase function" – a function that describes how it reflects light based on phase angle (sun-body-observer, $\varphi$), incidence angle (sun-body-normal, $i$), and emission angle (observer-body-normal, $e$). Usually with narrow-angle imaging, phase functions can be ignored for most planetary geology because those angles are nearly the same across the image. However, since MARCI images pole-to-pole, has an effective longitude width of ~30°, and includes pixels at 0° phase, the phase function cannot be ignored, otherwise brightness variations render the images more difficult to interpret. While that might be simple in theory, Mars' many different surface materials move with time, meaning that photometric corrections are difficult. This discussion also ignores scattering by Mars' atmosphere, which similarly affects images' appearance and makes blending multiple images more difficult.

# 3. Methods

The majority of this section is summarized in the flow chart in Fig. 1, and a pared down, worked example is in Appendix A. To streamline the text, some issues that require work-arounds with *ISIS* are described in Appendix B.

## 3.1. Input Dataset and Data Selection

MARCI Experiment Data Records (EDR) are distributed by NASA's Planetary Data System (PDS) through the Cartography and Imaging Sciences Node as 16-bit .IMG files; it should be noted that MARCI is natively an 11-bit instrument, with data compressed to 8-bit on-instrument, re-expanded to 11-bit on the ground, and then up-sampled to 16-bit for PDS delivery (Bell et al., 2009). The files are split into two, with the five VIS bands in one file and two UV in another due to the dramatically different file sizes (UV is *always* binned on-craft in 8×8, while VIS is roughly half the time taken in 1×1 and half in 2×2 binning or "spatial summing" mode). PDS metadata include notes about whether the data were "OK" or contain an "ERROR." However, as shown later, it is usually worth trying to process the ~6.5% of data which are ERROR because it can just mean a portion of the track is missing, while the rest is usable. Conversely, sometimes images noted as OK contain some odd artifacts that render them unusable. Unfortunately, due to the nature of MARCI images, it is almost impossible to screen until least some of the processing has been completed: if images fail any given processing step, that is a good indicator of a problem, but sometimes one needs to complete map-projection and even-odd merging before issues become apparent.

As of the time of this writing, nearly eight full Mars Years of data have been released to PDS (through PDS release 60, in March 2022). Using the Mars Year (MY) system specified by Cantor et al. (2010), MARCI began commissioning phase imaging at $L_s = 111.4°$ in Mars Year 28 (September 24, 2006), and data are released through $L_s = 93.1°$ in Mars Year 36 (August 31, 2021).

MARCI imaging has not been continuous, though it is still reasonably complete. Figure 2 shows a tick mark for every color that MARCI has taken at every $L_s$ for every MY that have been released to PDS. At the resolution of a journal article figure, the smallest visible gap is about 1–2 days. The largest gap was during the second year, when roughly three full Earth months of data were not returned during the 2008 global dust storm. The second-largest gap is a little over one Earth month during MY31, encompassing another large dust event. Additionally, MARCI-VIS is sometimes operated (or data transmitted) in a reduced capacity, where usually the two longest wavelengths (red and near-IR) are not returned. This shows up in Fig. 2 as those bands of color missing. During the commissioning phase (images that start with a "T"), alternative reduced modes that alternatively cut out orange and near-IR or orange and red were sometimes used, but generally those modes have not been used since.

### 3.2. Initial Image Processing

Initial processing is handled through the USGS's *ISIS* software:

- MARCI2ISIS: Ingests PDS-released data into the *ISIS* format. This step changes the file format, increases the bit depth from 16 to 32, and it doubles the images by splitting even from odd framelets to prevent *ISIS* from needing to handle multiple pixels in the same image in the same location. This step means the disk space needed is now ≈4× the original file size.
- SPICEINIT: Attaches SPICE data to the image that tells software how the spacecraft and camera system were oriented and moved during exposure, among other things.
- MARCICAL: Applies radiometric calibration specified by Bell *et al.* (2009), including dark-current subtraction, and application of the latest flat fields (at the time of this writing, those are from February 2020).

At each stage, it is possible that processing will fail, so one must monitor the output after each step and remove failed files from further processing. For example, one can read through the *ISIS* plain-text output file and look for errors, and one could run the *ISIS* STATS program to determine if at least ~40% of the pixels have data (due to slews and the even/odd framelet split, typically only ~40% of pixels in good images have valid data).

After MARCICAL, the *ISIS* .cub files from the VIS images will contain up to five layers of data, one for each color, while the UV images will contain two layers of data, one for each UV band. To process further, it is generally a good idea to split these into individual files. The *ISIS* EXPLODE program will split a .cub into its different layers. Next, one should re-name the files because of the occasional reduced-color mode of the VIS component. *I.e.*, one cannot always rely on the fourth band to be the red one, for it might be near-IR, or it might not exist at all. The *ISIS* command GETKEY can be used to query each .cub file to return what color is contained, and then simple Unix or other scripting languages can be used to uniformly re-name files with the proper color in the file name for easy reference later. It is common in MARCI literature where this is explicitly mentioned to name the VIS bands 1–5, in order of increasing wavelength, and UV bands 6–7, in order of increasing wavelength. I, personally, prefer to name them all in increasing wavelength order: 1–2 for UV and 3–7 for VIS.

### 3.3. Photometric Correction

Materials reflect different amounts of light towards an observer based on three primary angles: Phase ($\varphi$), incidence ($i$), and emission ($e$). If that reflectivity is properly modeled, and that model removed from the spacecraft image, then all that is left will be true reflectivity differences of the material, such as showing the real difference between a dusty versus a wind-cleaned surface on Mars.

*ISIS* has a comprehensive program, PHOTOMET, to apply several different photometric models, with or without an atmosphere, and with or without topography in the model. There are entire textbooks, Ph.D. dissertations, and innumerable research papers on deriving photometric properties using different models, so this discussion is not intended to do the field justice. To generalize for purposes of processing MARCI data, *some* photometric model should be used to eliminate as much of Mars' photometric effect as possible. This becomes a balance between how much effort one is willing to put into the photometric modeling versus how accurate they want the end result to be. In the interest of a streamlined discussion, Appendix C discusses some possible workflows as well as a less theoretical and more empirical flat-fielding approach done after more proper photometric modeling. Figure 3 illustrates the effects of photometric modeling on Mars where one set of photometric parameters is used: Due to Mars' changing surface and different materials, there will almost always be some residual effects which, in a color composite, manifest as rainbows because photometry is wavelength-dependent.

### 3.4. Cropping the Data

This next step can be omitted, but not doing this step will dramatically increase processing time: Crop the image width via *ISIS*'s CROP program. One would normally want to use *all* data available, especially with an instrument that does not *quite* provide continuous coverage from image-to-image during its daily swaths if there are any slews. However, there are two issues with pixels near the edge.

First, Mars does not fill the entire 1024-pixel-width of the detector. Instead, *MRO* keeps Mars within a certain buffer on either side of the MARCI detector, with thruster firings *very* roughly every 40° of latitude due to drifting (this is not true when there were intentional off-axis slews for targets of interest for other instruments). Therefore, in most images, the edges of the 1024-pixel image are empty space and processing them further would be pointless. Not only that, but unless removed, *ISIS*'s map-projection program CAM2MAP will take substantially longer to run because it must check and try to handle pixels that do not project onto the planet's disk.

Second, as noted earlier, one of MARCI's complications is an extreme fisheye lens, producing significant distortions near the edge of the field-of-view, for which the camera models in several bands could be improved. Even if the distortions were perfectly modeled, they image Mars at extremely high emission angles, much worse pixel scale, and through a lot of atmosphere, rendering surface features practically unrecognizable, except perhaps vague reflectivity differences. However, preserving some of those pixels might be preferrable to having gores – this is something the end-user will need to decide for themselves.

To implement a crop, one simply invokes *ISIS*'s CROP command, setting the pixel one wants to start on across-track, and the number of pixels to include from that point across. Further discussion of cropping and its effects on the time that map-projection will take is in Appendix D. Figure 3's panels (a) and (b) show an adaptive crop (Appendix D) with a minimum 80 or 96 pixels, and it shows close-ups in panels (c) and (d) illustrating the quality of those preserved versus excluded pixels. The purpose is to illustrate that, while less aggressive cropping will produce a more continuous-appearing daily mosaic, the information in those pixels has limited utility.

### 3.5. Map-Projection and Mosaicking

With calibrated images, one can map-project the data. This is accomplished through the standard *ISIS* program CAM2MAP. To make full use of the MARCI data, one can map-project the UV images at 12 pix/deg (ppd) and the VIS data at 96 ppd (4.936 km/pix and 617 m/pix at the center of the projection, respectively) – both are chosen to be evenly divisible, though slight differences would work. A complication is the spatial summing of 2 data, which on the average would be oversampled at better than ≈48 ppd. One must decide whether they want to over-sample half the data, under-sample half the data, or (depending on the application) under-sample all the data and produce more coarse results that are still useful for their research (Appendix A).

The time to map-project is extremely variable with MARCI images and is also highly dependent on the approach to cropping, described in the previous section and Appendix D. Appendix D gives more details, but in brief, on a Late-2020 MacMini, with a constant 96-pixel crop, images in Fig. 3 took an average of 12 minutes to map-project, but they ranged 0.3–49.7 minutes. In comparison, a no-crop run was stopped after 8 hours with no images having completed CAM2MAP. The reader is encouraged to perform their own experimentation to determine how many pixels are worth preserving, balancing their usefulness against the time to process them.

Once images are map-projected, they can be mosaicked together, and finally the even-odd framelets can be combined to create a gapless image (except for the first few months of data, T01–T02, and P01–P03, where the exposure cadence was too slow relative to the craft's motion and so gaps remain). *ISIS*'s AUTOMOS function can be used to do the merge. Setting the blending mode to "average" will ensure a relatively smooth transition between framelets, though in some cases (especially bluer wavelengths), the transition will still be visible.

Finally, if the desire is to not process one MARCI image but several, such as to build up one day, one month, or some other time period, then after the even-odd merging one can again use AUTOMOS to merge multiple images. Again, the recommendation here is to use the "average" blending mode to help reduce discontinuities and average out remaining, non-linear photometric residuals, such as the opposition surge on non-typical terrain.

As a first note, if one is interested in calculating backplanes, such as maps of the phase angle or pixel scale, *ISIS*'s PHOCUBE can be used. I have found that for the global strips, it is significantly faster to calculate these before CAM2MAP, and then mosaic them just as if they were the image, rather than calculate them afterwards from the map-projected image. If only studying small, local areas up to a few tens of degrees across, PHOTOMET is faster after running CAM2MAP.

As a second note, returning to required disk space, full MARCI images projected at 96 ppd in equirectangular coordinates are ≈2.2 GB on-disk, per color (and per-even/-odd band, until merged). Adding in backplanes increases storage requirements linearly. Without backplanes, for a 5-color VIS image at 96 ppd, the required disk space will go from 110 MB downloaded from PDS to ≈11 GB map-projected, plus at least twice that for scratch space when getting to the final map-projected data. This factor of ≈100 increase is one of the main reasons I think that MARCI data are hard to work with.

As a third note, the above simple average mosaicking tends to leave artifacts due to the extreme viewing geometry changes over a single MARCI frame. Both this author and others (*e.g.*, Wang et al., 2018) have found that more complex, weighted averaging produces a much smoother mosaic. For example, weighting by some form of $e^{-1}$ to reduce the impact of pixels at high $e$ can improve clarity where other images with smaller $e$ pixels exist at that location. Depending on the application, the reader may find this is preferrable. *ISIS* does not support this type of mosaicking, so alternative software must be used, such as Python's NumPy.

### 3.6. Color Composites

MARCI is inherently a color camera, so creating color composites is a natural final step after individual images for each color are processed. Creating color composites can also help one to visually identify changes in the surface, such as reddening or bluing, identifying clouds, and monitoring white seasonal frost against a rusty colored surface. There are innumerable ways to create color composites and different ways to mix-and-match the different MARCI bands. The most basic method is to simply read in three of the different colored images, and then output one as blue, one as green, and one as red in a three-color RGB image. Depending on what kind of consistency one wants throughout the mission, the three bluest VIS wavelengths might be preferred because those were used most often (blue = violet/blue, green = green, red = yellow/orange). Or, one might desire more contrast in wavelengths, and so one could use the reddest band with the violet and green when it is available, and change it to something else when those data do not exist.

Additionally, one will need to decide if further manipulation of the brightness values might be desired, purely for purposes of the color composites and visually identifying differences (clearly, manipulating these data, especially in a non-linear way, would not be desirable for a science application). For example, Mars is so faint in blue light that a linear RGB composite would be almost entirely red; therefore, boosting the value of the blue data is a reasonable manipulation for the composite, so long as it remains clear that has been done.

## 4. Sample Data

In this section, a few samples have been made to illustrate the quality of the imaging. The reader can see some other examples in earlier figures in this work, such as panels (e)–(h) in Fig. 3.

First is a small, 10°×10° area around Garni crater in Valles Marineris. Figure 4 shows six images of the same region. All panels are color composites, made the same way with the exact same scaling from the raw data, so they show real differences in color balance and brightness (provided the caveat that there were no gain changes that were unaccounted for). The top row shows different Mars Years and months to illustrate how the surface changes (the brightness difference in panel (c) is likely real, for the full dataset was processed and the dimming near this month of the year persisted across all eight Mars Years). The bottom row shows individual images, demonstrating changing reflectivity and weather patterns that are readily apparent in the MARCI data.

Second (Fig. 5) is a global mosaic of short-UV (260 nm) covering approximately one Mars Week (~4° $L_s$) soon after the northern summer solstice. The Figure's four panels show the effects of simple versus weighted averaging, where the weighted version clearly brings out more detail because it de-emphasizes pixels that are much more smeared (weighted here is based on a combination of emission and incidence angles). Clearly visible are some residual artifacts from the extremely low signal-to-noise of short-UV (average non-ice surface reflectivity is ~1–2%), but this processing clearly reveals delicate structures in the clouds, frost in Hellas crater, and the north polar cap after its summer solstice.

Third (Fig. 6) is a global mosaic of one Earth Month of averaged data, showing Mars averaged near its equinox going into southern summer ($L_s$ = 170.8°–188.2°). The brightness and color differences between most of Mars and the frosted-over southern polar areas is striking. Even in a one-month average, persistent cap clouds over Arsia Mons are visible, an optically thin haze over the north polar area is present, and striking surface patterns are clear. The faint green banding in Hellas crater is due to changing green-band reflectivity during the month.

The fourth example (Fig. 7) is a north polar projection showing one averaged Mars week near its northern summer solstice. This composite uses a different color mapping to emphasize that different information can be gleaned using different bands. In this case, the color is chosen to emphasize the clouds that are much more reflective in short wavelengths that show up as blueish throughout the composite, and also show the differences in reflectivity between the longest two bands, centered at 650 and 720 nm: The gradations between red, orange, yellow, and green are gradations in relative reflectivity where redder means the surface is more reflective in 720 nm, and greener more reflective in 650 nm. Besides the color, highly detailed structure is visible even at MARCI scales, and even the scales of this printed article (~5 km/pix).

Meanwhile, several authors have published peer-reviewed studies that are focused on using MARCI data. Those that cite the instrument papers include:

- Overviews or general observations: Malin *et al*. (2008), Rossi & van Gasselt (2010).
- Surface change detection or study: Geissler *et al*. (2016), Lujendra *et al*. (2017), Wellington and Bell (2020).
- Polar cap weather and surface features: Cantor *et al*. (2010), Calvin *et al*. (2015), Brown *et al*. (2016), James and Wolff (2018).
- Characterizing weather at landing sites: Tamppari *et al*. (2008, 2010), Holstein-Rathlou *et al*. (2010), Yao *et al*. (2020), Grant *et al*. (2018).
- Clouds: Clancy *et al*. (2008), Sánchez-Lavega *et al*. (2018), Wolff *et al*. (2019), Haberle *et al*. (2020), Kahre *et al*. (2020), Cooper *et al*. (2020), Guha *et al*. (2021), Wang & Abad (2021), Clancy *et al*. (2021).
- Dust storms: Wang and Richardson (2015), Guzewich *et al*. (2017), Heavens (2017), Cantor *et al*. (2019), Heavens *et al*. (2019), Ordonez-Etxeberria *et al*. (2020), Guha & Panda (2021).
- Atmospheric aerosols: Wolff *et al*. (2010), Clancy *et al*. (2016, 2019).

It is unusual that one can cite in a few lines most of the science papers that focus on using data from a NASA planetary camera that has been in operation for more than 15 years. While numerous first authors are listed, several common names are in the group, or they are listed among the second or other co-authors in the full papers, indicating even fewer research groups use MARCI than even this short list suggests.

## 5. Summary and Discussion

MARCI data represent an extremely rich image set spanning nearly eight Mars Years and seven different color bands with up to ≈0.5 km/pix scale. Despite the quality and quantity of the data, the number of papers that use it for science investigations is less than 50. The instrument's design and some inefficiencies and issues in commonly used processing software contribute to a lack of use and even a lack of understanding among many in the Mars science community about what MARCI is really capable of showing. It is hoped that this work, which presents a workflow on how to process MARCI data, might further MARCI's use.

Acknowledgements: This work was funded in part through internal awards by Southwest Research Institute, and a significant part through personal "free time." The author thanks the *ISIS*, NumPy, and GDAL contributors for their work developing, maintaining, and enhancing the programs used that allow the workflow described in this work. The author thanks M. Wolff and an anonymous reviewer for their helpful reviews, and B.P. Gisclair and D.E. Stillman for additional text comments.

Conflict of interest: The author declares no conflicts of interest relevant to this study; the author is not a member of the MARCI nor *MRO* teams nor employed by Malin Space Science Systems (MSSS).

Data Availability Statement: *ISIS* processing software is freely available from the United States Geologic Survey (https://isis.astrogeology.usgs.gov). MARCI data are available from NASA's PDS's Cartography and Imaging Science Node at https://pds-imaging.jpl.nasa.gov/volumes/mro.html. The code implementing the workflow described in this work is not publicly accessible due to intellectual property restrictions, though a sample, basic workflow is explicitly given in Appendix A. However, the Southwest Research Institute welcomes inquiries regarding licensing, non-disclosure agreements, or access/use in collaborative or cooperative agreements (please contact the corresponding author).

## Appendix A. MARCI Processing Recipe with *ISIS*

This Appendix provides a basic *ISIS* processing pipeline implementing most of the steps described in this work needed to successfully process a MARCI image. Various optimizations and photometric corrections are not included in the interest of ease-of-use and understanding. This section assumes that the user has already successfully installed *ISIS* and has it running properly, and it is written for *ISIS* versions 3.9–6.0.0; there is no guarantee that it will work with earlier nor later versions. It also assumes the user has downloaded a processable MARCI-VIS file, which will be called "image" here. Every line with a # before it is a comment, every line with a > before it is a command. The image used to validate this code is N18_069581_0524_MA_00N113W.IMG, available at the time of this writing from https://pds-imaging.jpl.nasa.gov/data/mro/mars_reconnaissance_orbiter/marci/mrom_1319/data/

```
#Import MARCI data from PDS into ISIS format.
#The "flip=no" is required unless a certain bug has been fixed.
#Additionally, there might be a "NoExposureTimeDataFound" warning when
running this command; it can be ignored.
> marci2isis from=image.IMG to=image.cub flip=no

#Attach SPICE data.  Note that the output of marci2isis will have created two
files, an even and odd, which must be processed separately from here on.  The
"cknadir=true" is required unless another bug has been fixed.
> spiceinit from=image.odd.cub cknadir=true
> spiceinit from=image.even.cub cknadir=true

#Basic calibration.
> marcical from=image.odd.cub to=image.odd.l1.cub
> marcical from=image.even.cub to=image.even.l1.cub

#Separate the cubes into different wavelength bands.
> explode from=image.odd.l1.cub to=image.odd.l1
> explode from=image.even.l1.cub to=image.even.l1

#The above will output images with "band0001" through "band0005" if your
image was visible-light and had all five colors.  In the workflow in the main
body, I describe using "getkey" to query the header information of each
output file to rename them into something consistent, per band.  That is not
shown here.

#The next step is to perform some sort of photometric correction.  That is
not shown here, but it is assumed that the files now have a ".l3" attached to
the file name to indicate photometric correction has been applied.  To work
this example on your own, simply rename the files after the "explode" to have
```

a ".l3" before the ".cub" and remove the ".l1".

```
#Crop the data, if you desire.  It is assumed here that your image is in 1×1
spatial summing mode, so a 108-pixel crop on both sides removes ≈20% of the
image width for an image 808 pixels wide; an aggressive crop is applied here
to speed CAM2MAP for this worked example.  These lines would need to be
repeated for each band number, but in this example I have selected band 5
("red/near-IR").
> crop from=image.odd.band0005.l3.cub to=image.odd.band0005.l3.crop.cub
sample=108 nsamples=808
> crop from=image.even.band0005.l3.cub to=image.even.band0005.l3.crop.cub
sample=108 nsamples=808

#Create a map-projection template.  Users will often have this already made
and "canned" to refer to, but I explicitly create it here since it is
required.  I have selected a very low-resolution 24 ppd to save time.
> maptemplate map=AA_equ_24ppd.map projection=equirectangular clon=0 clat=0
targopt=user targetname=Mars londom=180 rngopt=user minlat=-90 maxlat=90
minlon=-180 maxlon=180 resopt=ppd resolution=24

#Map-project.  This would again need to be repeated for each band; in this
example I have selected band 5.  This can take a long time, so wait at least
a half hour before worrying.  If you use the image noted at the top of this
Appendix, the maptemplate above, and the cropping above, it should take just
a minute or so (test performed on a 2017 MacBook Pro).
> cam2map from=image.odd.band0005.l3.crop.cub to=image.odd.band0005.l2.cub
map=AA_equ_24ppd.map pixres=map defaultrange=map
> cam2map from=image.even.band0005.l3.crop.cub to=image.even.band0005.l2.cub
map=AA_equ_24ppd.map pixres=map defaultrange=map

#Merge the even and odd map-projected files.  Again, repeat for each band.
> ls image.*.band0005.l2.cub > merge.lis
> automos from=merge.lis mosaic=image.band0005.l2.cub priority=average

#Extract just the image data (because "priority=average" in AUTOMOS, the
output cube will have two bands, one being the image data and the other being
the counts per pixel).  Again, repeat for each band.
> explode from=image.band0005.l2.cub to=image.band0005.l2
> rm image.band0005.l2.band0002.cub
> mv image.band0005.l2.band0001.cub image.band0005.l2.cub

#Create a simple image that can be read in a standard image viewer, if
desired.  Again, repeat for each band.
> isis2std from=image.band0005.l2.cub to=image.band0005.l2.png
```

## Appendix B. Known *ISIS* Quirks, Issues, and Bugs as of version 6.0.0

This Appendix is intended to briefly describe some of the issues working with MARCI data using *ISIS* v6.0.0. I must preface this by saying that I appreciate the immense amount of work the developers put into creating and maintaining *ISIS*, but it is a simple fact that there are some issues with *ISIS* when processing MARCI data which can hinder success.

1. MARCI2ISIS: MARCI can scan across Mars either North-to-South or vice-versa, but the detector is exposed the same way regardless. This can mean that the framelets might need to be flipped in order to align one with the next properly. However, the automatic setting in at least some *ISIS* versions can lead to bugs later, so I have found that explicitly setting "flip=no" in the command of MARCI2ISIS, regardless of what image one is processing, will mitigate this issue.

2. SPICEINIT: Setting "cknadir=true" is required for processing MARCI, which forces *ISIS* to calculate nadir pointing data.

3. MARCICAL: As of the time of this writing, there is an open question about whether MARCICAL is applying calibrations correctly, especially the flat fields.

4. CAM2MAP: CAM2MAP does not handle the "odd" MARCI framelet images well *if* they have been subjected to the kind of cropping or sub-imaging described in Appendix D (as of the time of this writing, *ISIS* developers are working on this problem, so readers are encouraged to check the latest version and documentation). While there are alternative map-projection methods in *ISIS*, such as PIXEL2MAP, PIXEL2MAP produces its own artifacts when used in default mode, or it quits and reports errors when used with more customized parameters for MARCI data. At present, there is not a good work-around for these issues other than experimentation with different crops and determining if the artifacts manifest or not, where less aggressive cropping means they are less likely, but the data will take longer to map-project.

5. Processing Backplanes: Nominally, one would calculate backplanes and they would all be in one *ISIS* .cub file. One can then map-project that file, and all backplanes would be in the output. Unfortunately, while this works with linescan and framing cameras, it does not work for push-frame, where the multiple backplanes do not propagate. As of *ISIS* version 6.0.0, developers are working on this issue, so the reader is encouraged to check the documentation to see if it has been fixed. If not, the different backplanes must be split using EXPLODE before CAM2MAP, and CAM2MAP run separately on each one. However, there is one time-saver here: PHOCUBE will calculate the backplane on an even or odd framelet image as though the framelets are not separated. That means PHOCUBE need only be run once per even/odd pair, and each backplane map-projected only once per image, saving both processing time and disk space.

## Appendix C. Photometric Correction Discussion

This Appendix is a further discussion of the main text's section 3.3 on photometric modeling.

On an airless body with just one surface material, the process of photometric modeling is generally straightforward: Select images of the surface – ideally of the same location – that provide a range of $\varphi$ (phase), $e$ (emission), and $i$ (incidence), and then run an optimization code of some sort that iterates through possible values for the various parameters of whatever photometric function is being used. Typically, the goal is to subtract the value modeled from the value observed and minimize the root-mean-squared (RMS) of that difference (RMS is preferred over standard deviation, for the goal is to get close to 0, while the standard deviation has the mean removed so will always be centered around zero). The parameter space search can take a lot of computer time because of the numerous parameters and possible values, but it is nonetheless straightforward. However, it usually should have a human in the loop in order to ensure the minimum is a reasonable model, since most photometric parameters interact with each other in non-linear ways.

For Mars, photometric modeling is not as straightforward because Mars has many different surface materials; those materials move; and Mars' atmosphere varies in thickness, opacity, and extinction properties as a function of wavelength. Therefore, unless one only wants to examine a small area of Mars, one will either need to create *many* different photometric parameter sets and create a sort of blended photometric correction across Mars, for relatively short intervals of time, or one will need to perform some sort of average across Mars' surface and across some time interval. That average will mean some effects are not wholly removed in some areas of the planet because the photometry is not correct at those locations. Additionally, one must check or curate the input data before running the photometric parameter optimization code because Mars is capricious: The very anomalous features that often make Mars interesting will throw off the photometry results because they do not correspond to the more common surface reflectivity properties (*e.g.*, dust or cloud events). Therefore, this process cannot be fully automated easily. Finally, the photometry must be derived for each MARCI color that is going to be processed, for photometric properties are dependent on wavelength. PHOTOMET will either calculate $\varphi$, $e$, and $i$ backplanes and discard them, or you can calculate them, use them as input to PHOTOMET, and save them for later.

As a general recommendation, I have found that sampling several sites across Mars in monthly intervals *tend* to provide reasonable results with neither too much manual effort nor time. I use the Hapke-Henyey-Greenstein photometric model without an atmosphere because the atmosphere adds significantly more complication, and I have found reasonable results can be derived for most of the usable MARCI imaging area without it (see various Figures throughout this work). For a renormalization value (so the mean is not 0), also accomplished in *ISIS*'s PHOTOMET, I use the original photometry input data, averaging those image values when $\varphi$ = 30–60°, $i$ < 60°, and $e$ < 30°, and I have found that to work reasonably well. It should be noted that when very different photometric materials are in the same image, this can result in clear issues. For example, Fig. 3 has the photometric parameters optimized for the majority of the dusty surface, but the aphelion cloud belt, north polar hood, and south polar frost have different photometric properties which cause the mosaic to look poorly composed.

After applying the above theoretical correction, a final practical correction of dividing out empirical flat fields can be applied for two primary reasons: First, there could be some remaining, across-track photometric residuals, despite the modeling, that were not removed in PHOTOMET (usually due to the atmosphere not being modeled and removed). Second, the currently available flat fields in *ISIS* either do not fully remove the sensor pattern noise, or the *ISIS* program MARCICAL does not correctly apply them (see Appendix B). This step does not need to be done, but it produces a better end-product at the pixel level. I have found it particularly useful for removing extra sensor pattern noise when using full-resolution MARCI data to analyze features at a local, roughly kilometer scale. While concern might arise that this will destroy the photometric validity of the data, I have found that pixel values overall are not significantly changed in this process, but pattern noise is substantially reduced, especially in the UV and violet. It is also true that the flat-field process is usually applied before photometry; however, these are *empirical* flats, so if they were done before photometric modeling, they would remove some of the photometric effects, and then throw off the photometric modeling. Since the primary goal of this step is to further remove pattern noise from the sensor and optics, doing it at this stage is appropriate.

To create custom flat fields, the framelets from each image must be combined separately for the seven colors, separately for each spatial summing mode. This cannot be done in *ISIS*, so instead I use GDAL and NumPy, free Python packages. Each framelet is stacked, and the median value of that stack is taken, producing one master framelet flat. The median of the inner portion is taken (to prevent the atmosphere from affecting the results), and the master framelet flat is divided by that median value so that the median of the new flat is 1.0 (except the edges). Then, because each MARCI image can start at a different row in the final image (and certainly the even-odd splits will start at different rows), a custom full flat must be made for each image. This is also accomplished in GDAL by determining the row at which data begin, how long the MARCI image is, and then tiling the flat framelet down the length of the new flat image. *ISIS*'s RATIO program can then be used to divide the image by the flat, providing the final calibrated image.

## Appendix D. Cropping Discussion

This Appendix is a further discussion of the main text's section 3.4 on cropping and the effects of cropping on the time that map-projection will take.

One can approach cropping in two different ways. First, one could do a simple crop on both sides of the image, attempting to remove the majority of off-planet pixels while preserving most useful data. I have found a crop of 96 pixels for a spatial summing of 1 to be a reasonable balance between those competing factors, and more aggressive cropping did not significantly speed map-projection (96 pixels is evenly divisible by 8, so the same relative crop can be scaled to any of the spatial summing modes). Second, one could attempt an adaptive crop, parsing through the image and dividing it into groups of framelets that have the same number of columns on Mars, crop the image into those groups, and then process each group separately through the next several steps in the main text, merging at the end.

For an idea of the time required to map-project based on different cropping schemes, my test computer was a Late-2020 MacMini. For a consistent 96-pixel crop (not illustrated), the images took an average of 12 minutes to map-project, but they ranged 0.3–49.7 minutes. The extreme range is due to extra, off-planet pixels in some images. With an adaptive algorithm that divided the images multiple times to remove sky pixels, while using at least a 96-pixel crop (Fig. 3b), it took about 5.7 minutes on average to map-project an image, and the range was 0.3–21.7 minutes, cutting in half the longest time with the fixed crop. Shrinking the minimum crop to 88 pixels raised the time to 10 minutes on average with a 0.5–25.7 minute range. Using the same adaptive algorithm with a minimum of an 80-pixel crop (Fig. 3a), map-projection took an average of 28 minutes per image with a range 10–49 minutes, indicating this is the approximate limit at which all images will contain at least a few sky pixels. Meanwhile, a non-adaptive, no-crop run was stopped after 8 hours with no images having completed CAM2MAP.

It must be noted that this test was conducted with *ISIS* 6.0.0, and future versions might optimize off-planet or extreme-*e* projection, so the user is encouraged to test different (or no) crop.

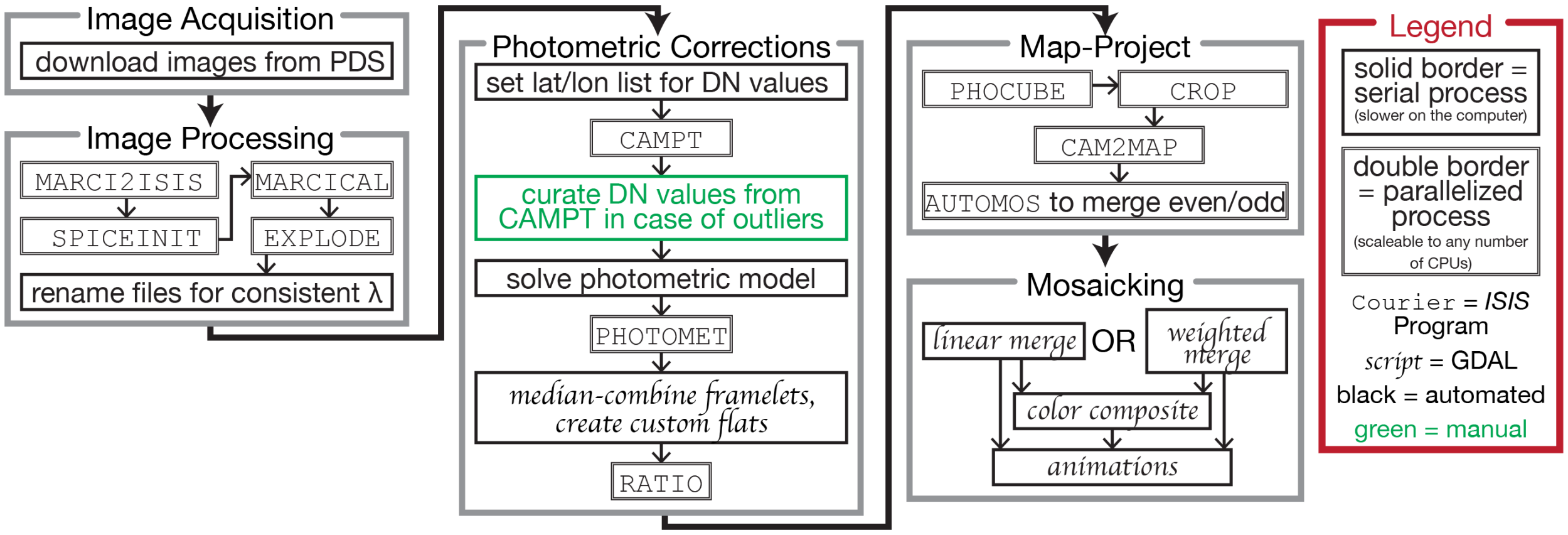

Figure 1. Flowchart illustrating the processing pipeline described in this work.

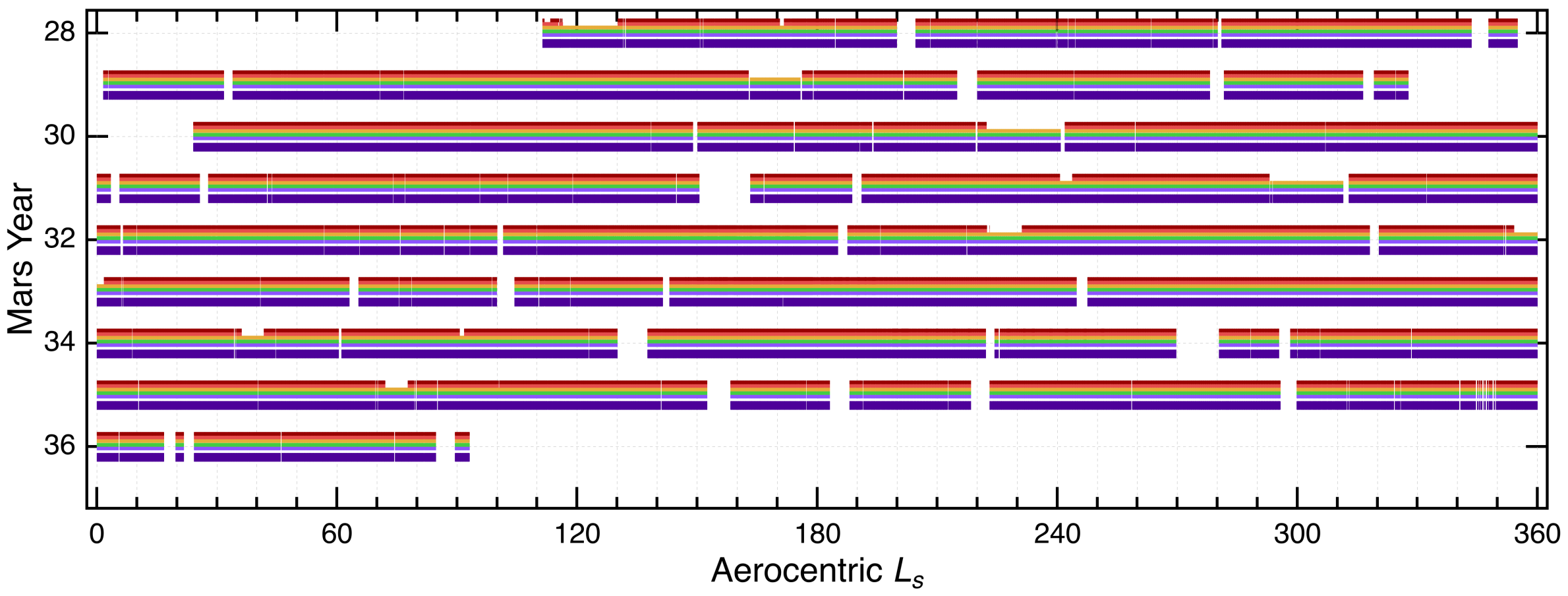

Figure 2. Illustration of MARCI time series for visible and UV, colored in bars to approximately match the human visual interpretation of the center of the bandpass (except UV and red/near-IR). There is one mark per observation in each color, and nominal temporal cadence was slightly less than 2 hours, but gaps $\lesssim$1 day do not appear at the presented resolution. The longest coverage gap is in the second year for a period of $\approx$3 Earth months (B12–15 images) during a global dust storm. MARCI can operate in a reduced color mode, most commonly omitting the longest two wavelengths (red, and red/near-IR), shown in this Figure as missing those colors.

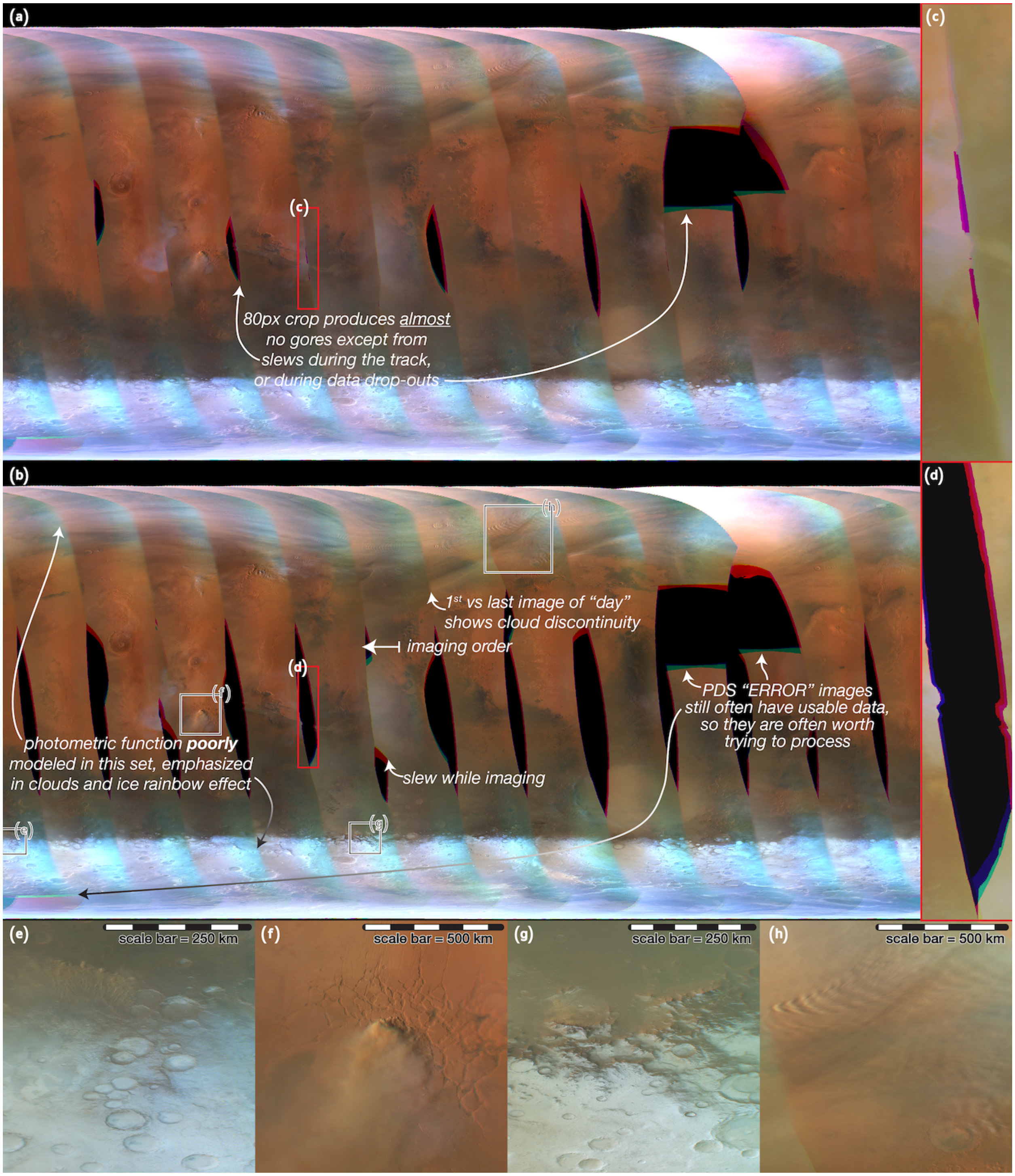

Figure 3. One Mars Day three-color composite mosaic (images G05_020361 through G05_020373; violet/green/NIR), in a ±180° longitude domain (all other global mosaics in this work are in this domain), illustrating several points. First, panels (a)–(b), with closeups in (c)–(d), show the difference between (a) cropping with an adaptive method that crops at least 80 pixels on both sides of the image, versus (b) cropping with an adaptive minimum of 96 pixels: While gaps are almost filled in (a), the data in those locations are fairly useless due to the geometric distortion, shown as (c) versus (d). Second, the color offsets at the edges, most easily visible in (d), are due to the different colors being exposed at different times as the full swath was built up. Third, those panels show the effects of using one set of photometric parameters optimized for the dusty regions, for they model neither the clouds nor ice well. Fourth, this Figure contains three MARCI images that are noted as "ERROR" in the PDS, but all three have useful data still contained in them. Fifth, this Figure demonstrates that even with the averaged photometry, a significant amount of useful information can be gleaned, as demonstrated in panels (e)–(h), which are identified with their locations in panel (b): (e) Local dust storm coming off the seasonal frost, (f) local dust storm casting a shadow with some water-ice to the lower-right (bluer in color), (g) detailed frost lines on anti-sun slopes, (h) gravity waves in polar hood clouds.

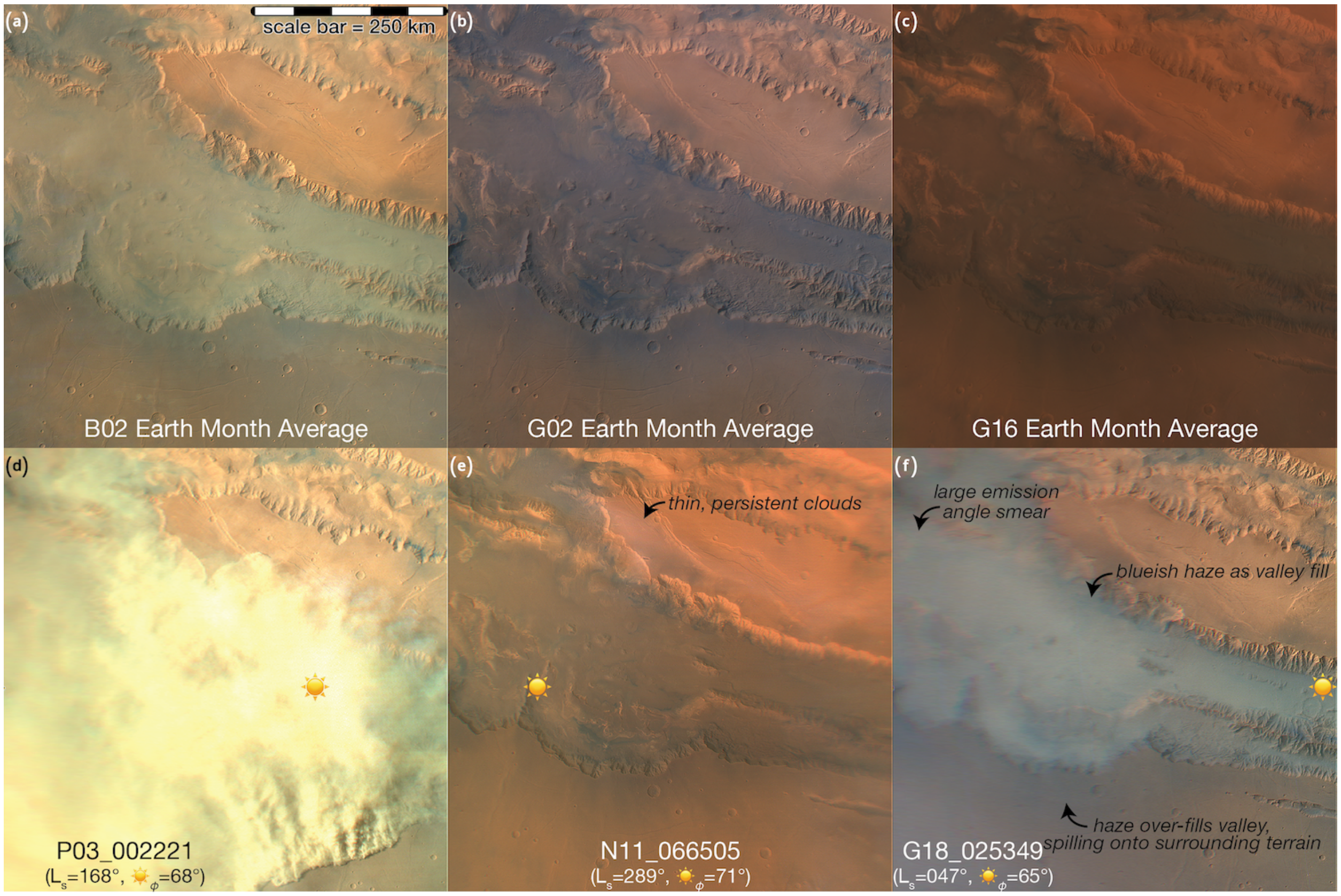

Figure 4. Three-color composites (violet/green/NIR) of the 10°×10° area around Garni crater in Valles Marineris (centered at –11.5°N, –69.7°E). Panels (a)–(c) show one Earth month averages during different $L_s$ points of different Mars Years, illustrating color, brightness, and feature changes between years and months. Panels (d)–(f) show selected one-day color composites illustrating various weather phenomena clearly visible at MARCI pixel scales. Scale bar is at the top of panel (a), and the sun icon in the bottom row indicates the approximate sub-solar longitude of the image, also stated as ☀$_\phi$ in the bottom text.

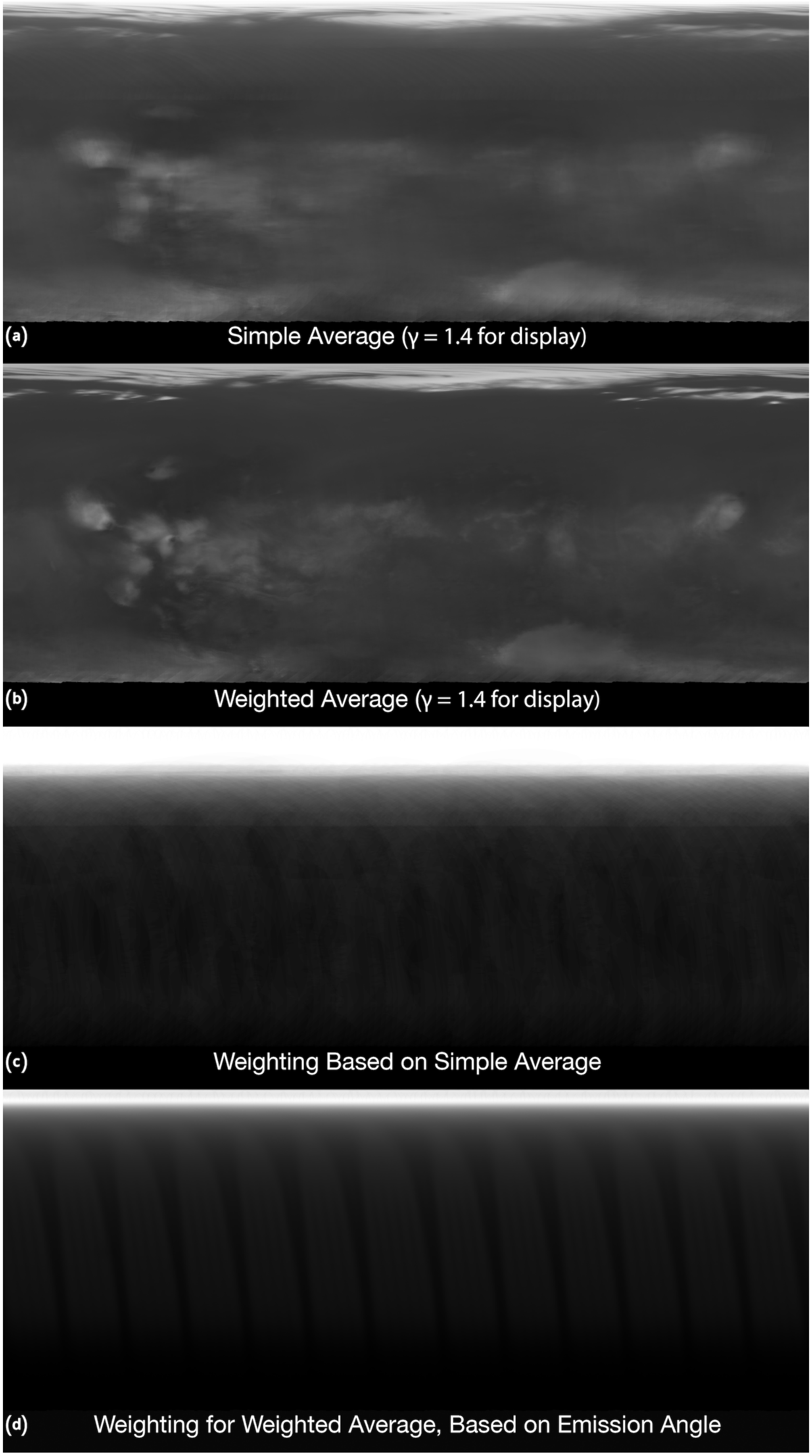
(a)
Simple Average (γ = 1.4 for display)
(b)
Weighted Average (γ = 1.4 for display)
(c)
Weighting Based on Simple Average
(d)
Weighting for Weighted Average, Based on Emission Angle

Figure 5. (a)–(b) Ten-day mosaic of the short-UV (~260 nm) color (images N01_062577 through N01_062713), where (a) is a simple average of the data, while (b) is a weighted average; panels (c) and (d) show the corresponding weights for the simple vs weighted averages, where the simple is just the counts. A gamma ($\gamma$) adjustment of 1.4 has been applied to (a)–(b) to bring out detail in darker areas for display ($DN_{new} = max_value \cdot (DN_{old} \div max_value)^{(1 \div \gamma)}$, where max_value is $2^8$–1 for an 8-bit image, or $2^{16}$–1 for a 16-bit image). The difference in the top two images shows the power of judicious weighting based on where on the detector MARCI data are best. There are clearly some remaining artifacts in this low-S/N wavelength range (which more careful photometry might improve), where any pattern at a ≈93° angle counter-clockwise to the equator must be treated with caution, and the bottom ~10° has photometric issues, but in general this mosaic shows superb detail given the quality of the original images, including significant structure throughout the aphelion cloud belt and volcanic lee clouds. Note: Since this image set is near the northern summer solstice, images have been cropped to extend to only 81°S to save space.

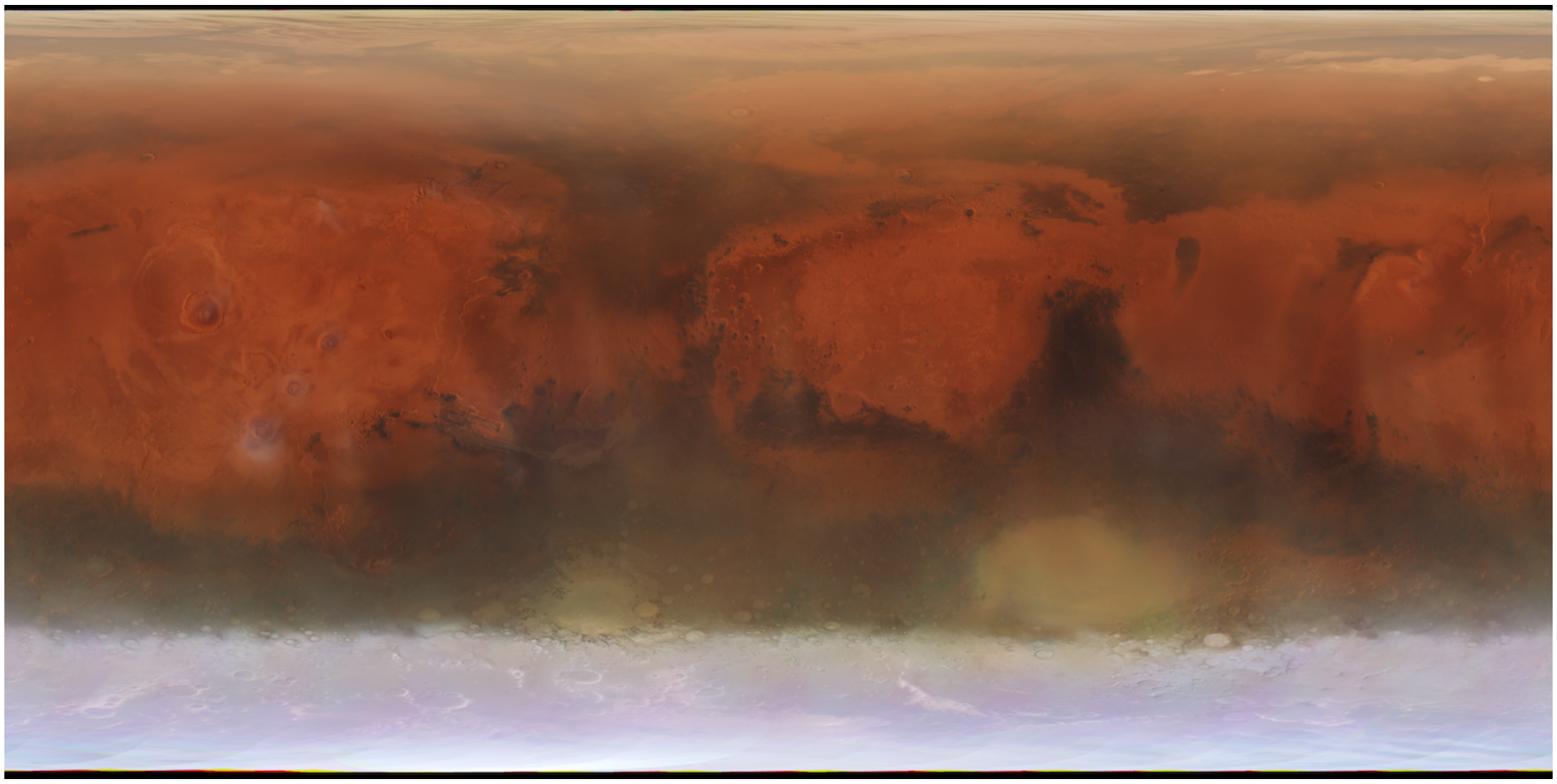

Figure 6. Three-color composite of one Earth month average near the equinox ($L_s$ = 171–188°; all F05 images; violet/green/NIR). Brightness values have been adjusted for better display of the large dynamic range for a journal article.

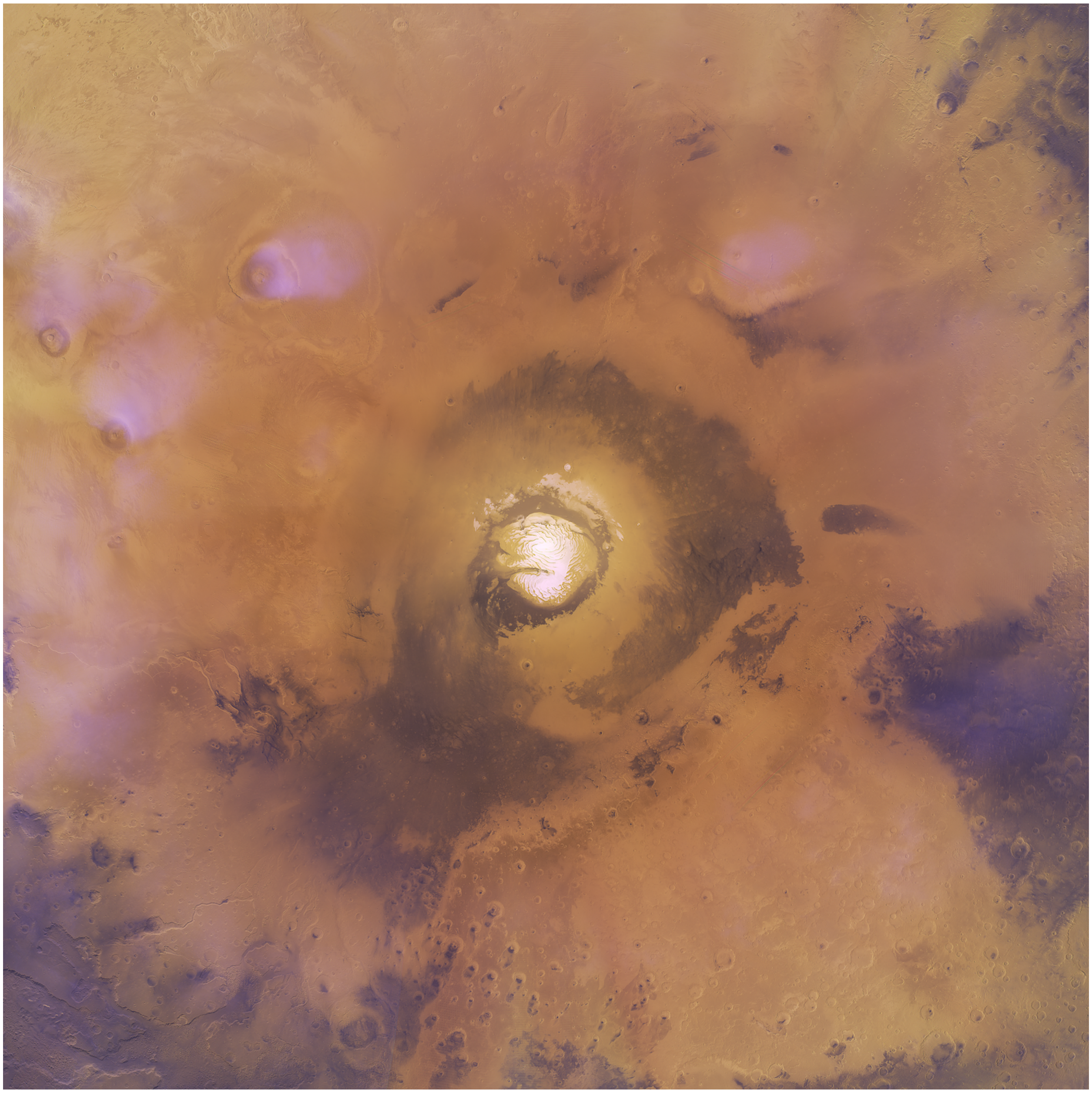

Figure 7. Three-color composite (violet/red/NIR) of ten days around the northern summer solstice in a north polar stereographic projection (images G21_026535 through G21_026665; center is +90°N, while the northern-most south latitude displayed is 0°, and the image extends farther south in the corners). Two small processing artifacts are visible as slight radial lines near the 12:30 position and 10:30 position, while four concentric bands of red/blue are visible near 1:00, 4:00, and 11:00 that correspond with data drop-outs.